\documentclass [twocolumn,prl,a4paper,altaffilletter,superscriptaddress,amsmath,amssymb]{revtex4}
\usepackage{times}
\usepackage{graphicx}
\usepackage{bm}
\begin {document}
\title{Practical vortex diodes from pinning enhanced YBa$_2$Cu$_3$O$_{7-\delta}$}
\author {S. A. Harrington}
\author {J. L. MacManus-Driscoll}
\author {J. H. Durrell}
\email{jhd25@cam.ac.uk}
\affiliation {Department of Materials Science and Metallurgy, University of Cambridge, Pembroke
Street, Cambridge, CB2 3QZ, UK.}

\begin {abstract}
We identify a scalable, practical route to fabricating a superconducting diode. The device relies for its function on the barrier to flux vortex entry being reduced at the substrate interface of a superconducting pinning enhanced YBa$_2$Cu$_3$O$_{7-\delta}$ nano-composite film. We show that these composite systems provide a practical route to fabricating a useful superconducting diode and demonstrate the rectification of an alternating current.
\end {abstract}
\maketitle

A Type-II superconducting system where vortices can move more easily forward than backwards constitutes a vortex diode. Such a system will give rise to a superconducting diode when a transport current is applied \cite{Krasnov1997}. This type of device exhibits no voltage drop across it until a critical current is reached and the magnitude of this critical current will depend on the direction of current flow, this is the opposite of the behaviour of a semiconductor diode. A practical implementation of such a device could find applications in AC-DC conversion, fault current limiters and superconducting electronics.

Vortex diodes have been demonstrated utilising ratchet effects from asymmetric pinning potentials \cite{Wordenweber2004,Lu2007,Shalom2005,VandeVondel2005,Yu2007} and asymmetric driving potentials\cite{Cole2006}. These studies have concentrated on the fundamental interest in these model physical systems and the effects demonstrated have not been compatible with large scale fabrication methods or high currents.  Additionally, several workers\cite{BROUSSARD1988,JIANG1994,JIANG1995}, notably Jiang \emph{et al.}, have reported superconducting diode-like characteristics in low-$T_{\textrm{c}}$ superconducting thin films due to surface pinning effects.

\begin{figure}
\includegraphics[width=9cm]{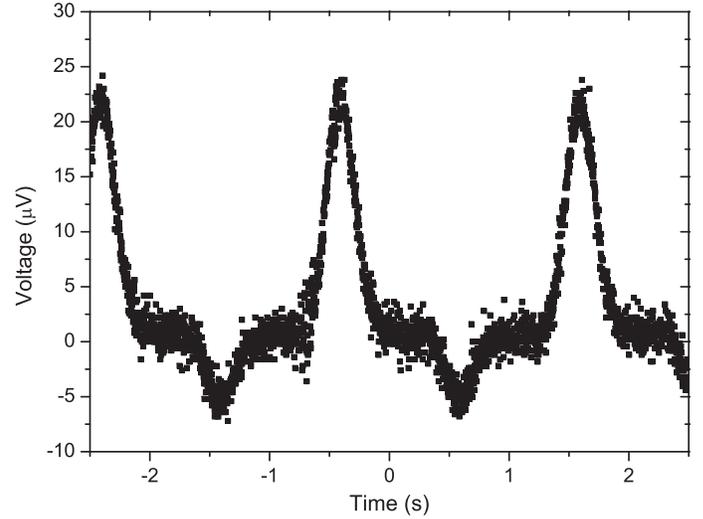}
\caption{\label{fig:4} Voltage measured across the measurement track when a 90 mA (peak-peak) 0.5 Hz sine wave was applied in a 0.5 T in-plane field.}
\end{figure}

In this letter, we describe the superconducting diode behaviour of the cuprate-superconductor/non-superconducting oxide nano-composites which have recently driven order of magnitude increases in the current carrying capacity of YBa$_2$Cu$_3$O$_{7-\delta}$\cite{Harrington2009, B.Maiorov2009,Mele2008}(YBCO). Figure \ref{fig:4} shows the rectification of a low-frequency AC signal using a sample investigated in this paper, a YBa$_2$Cu$_3$O$_{7-\delta}$ + RE$_3$TaO$_7$ film, a standard pinning enhanced cuprate material.

In YBa$_2$Cu$_3$O$_{7-\delta}$ one of the most prominent features in the dependence of $J_{\textrm{c}}$ on the orientation of magnetic field is the maximum found when the field is applied parallel to the \emph{a-b} planes. This maximum, often termed the `intrinsic' pinning peak, was discovered early in the exploitation of the cuprate superconductors\cite{roa90}. The peak is a result of a complex mixture of contributions including surface pinning, the Ginzburg-Landau anisotropy of the material\cite{civ04}, the layered structure of the material\cite{nis91,tac89} and in-plane oriented pinning centres\cite{Maiorov2007}. The importance of surface pinning is often neglected although prior work on vicinal films, in which geometrical effects are separated from intrinsic effects, shows that it can be of a magnitude equal to the intrinsic pinning \cite{ber96_3}.

\begin{figure}
\includegraphics[width=9cm]{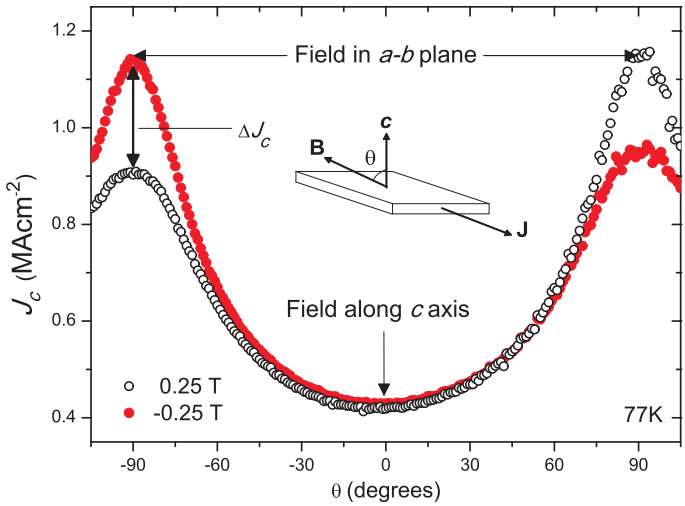}
\caption{\label{fig:1} Variation of $J_{\textrm{c}}$ with angle of applied field at +ve and -ve 0.25 T in a YBa$_2$Cu$_3$O$_{7-\delta}$ + 5 mol \% Yb$_3$TaO$_7$ film. The striking difference between the in-plane maxima when the Lorentz force direction is reversed is clearly apparent. The angle $\theta$ is that between the applied magnetic field and the $c$-axis with $\theta$=$\pm$90$^\circ$ representing a field laying in the plane of the film. The field was rotated in a plane perpendicular to the current direction.}
\end{figure}

We have experimentally observed, in several different types of pinning-enhanced YBa$_2$Cu$_3$O$_{7-\delta}$ [YBa$_2$Cu$_3$O$_{7-\delta}$ + BaZrO$_3$, YBa$_2$Cu$_3$O$_{7-\delta}$ + RE$_3$TaO$_7$ and RE$_a$RE$_b$Ba$_2$Cu$_3$O$_{7-\delta}$], that where the magnetic field angular dependence of $J_{\textrm{c}}$ was measured, maxima arising from opposite magnetic field directions can be significantly different. One peak characteristically exhibits a reduced maximum, as shown in figure \ref{fig:1}; a flattened maximum; or even a local minimum superimposed on the `intrinsic peak'. From the experimental geometry we have determined that the suppressed $J_{\textrm{c}}$ maximum corresponds to entry of vortices at the film-substrate interface, conversely the larger peak arises from entry of vortices at the free surface.

This effect arises for in-plane fields because the strong demagnetising factor associated with the large anisotropy of a thin film is eliminated when magnetic field is applied parallel to the film. In this geometry the difference between the surface barriers at the two interfaces becomes apparent. Wherever there is an abrupt change in superconducting properties over a short distance\cite{cam72} a barrier to vortex motion arises. The barriers to vortex motion at the surface of a superconductor are described by the Bean-Livingston model\cite{Bean1964}. Of the entry and escape barriers the entry barrier is larger\cite{clem74}, they are differently affected by surface imperfections. Vodolazov \emph{et al.}\cite{Vodolazov2005} note that an asymmetry in critical current for reversed Lorentz force will arise wherever surface pinning is significant. The asymmetry is most significant at small fields where the surface pinning barrier is large compared to the pinning force density on the vortex lattice. Kumar \emph{et al.} have experimentally demonstrated that the surface barrier can be greatly reduced by grading the superconductor interface\cite{KUMAR1994}.

\begin{figure}
\includegraphics[width=9cm]{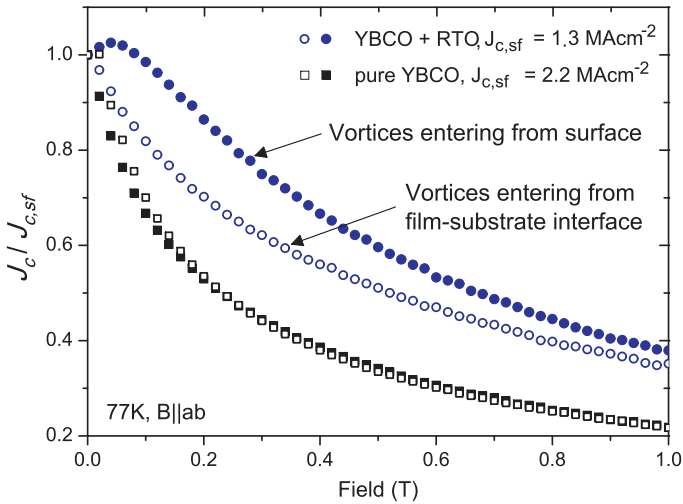}
\caption{\label{fig:2} Comparison of the change in $J_{\textrm{c}}$ , scaled with respect to the self-field $J_{\textrm{c,sf}}$, with reversed Lorentz force in pure and Yb$_3$TaO$_7$ containing YBa$_2$Cu$_3$O$_{7-\delta}$ at various fields. Closed symbols show behaviour when vortices enter at the free surface and open circles when vortices enter at the film-substrate interface.}
\end{figure}
On the assumption that the effect seen in our films was due to a graded interface reducing surface pinning at the film-substrate interface, a number of rare earth tantalate (RTO) doped films were prepared\cite{Harrington2009} in which growth conditions were chosen so as to encourage disorder at the film-substrate interface. This involved using fast deposition rates and low growth temperatures to slow the assembly of the pinning enhanced phase. The resulting films had normal transition temperatures ($>$ 91 K) but due to the non-optimum growth conditions somewhat reduced self-field critical currents compared to that of standard RTO + YBa$_2$Cu$_3$O$_{7-\delta}$ thin films. YBa$_2$Cu$_3$O$_{7-\delta}$ + 5 mol \% Yb$_3$TaO$_7$ composition films were found to give the largest anisotropy and it is data from such a film that we present in this letter. Measurements were performed using a two-axis goniometer mounted in an 8 T magnet\cite{her94}, on a 1 mm long 500 nm x 90 $\mu$m track patterned using standard photolithographic techniques. A four terminal measurement was used to obtain $IV$ characteristics from which critical current values were determined using a voltage criterion of 0.75 $\mu$V.

Figure \ref{fig:1} clearly shows the large difference in critical current when the Lorentz force is reversed, for an in-plane magnetic field. It was also found that a reversal of the measurement current direction produced the equivalent anisotropy.

In figure \ref{fig:2} we show the striking difference between the behaviour of the interface engineered YBa$_2$Cu$_3$O$_{7-\delta}$ + Yb$_3$TaO$_7$ film and a reference YBa$_2$Cu$_3$O$_{7-\delta}$ sample grown under standard conditions. It is apparent, that while the pure YBa$_2$Cu$_3$O$_{7-\delta}$ film does exhibit some surface pinning anisotropy the effect is opposite to, and much smaller than, that seen in the interface engineered film. We conclude that the small effect seen in the pure YBa$_2$Cu$_3$O$_{7-\delta}$ film is due to the free surface being slightly rougher than the film-substrate interface.

\begin{figure}
\includegraphics[width=9cm]{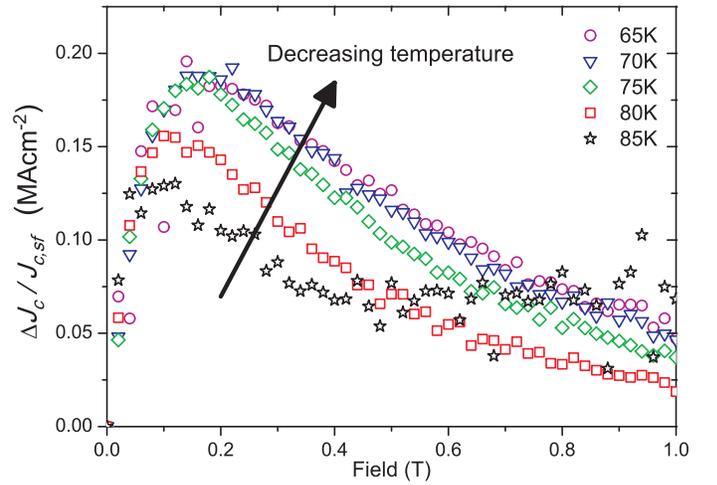}
\caption{\label{fig:3} Variation with field of the magnitude of the difference between the $J_c$ of the two intrinsic peaks ($\Delta J_{\textrm{c}}$) normalised to the self-field $J_{\textrm{c,sf}}$. This dependence is shown at various temperatures.}
\end{figure}

In figure \ref{fig:3} the magnitude of the difference in the surface pinning at the two interfaces increases with decreasing measurement temperature. The figure also highlights the decreasing anisotropy with increasing field for all temperatures measured. The anisotropy extends beyond 1 T with a peak being observed around 0.18 T.  An explanation involving surface pinning is consistent with the observed behaviour for several reasons: the effect disappears in high fields, consistent with the Bean-Livingston model\cite{Bean1964}, unlike vortex channelling effects which are field independent \cite{durrgb03,dur04}; no asymmetric bulk pinning is present in the films; and the films have large critical currents meaning that channelling at poor growth-grain boundaries is not likely. Further, the sample does not exhibit the \emph{c} axis peak in $J_{\textrm{c}}$ which occurs when nanorods are present\cite{Goyal2005a}, suggesting that channelling along nanorod arrays is unlikely to be present.

In conclusion we have demonstrated a superconducting diode with a 25\% variation in critical current in pinning enhanced YBa$_2$Cu$_3$O$_{7-\delta}$ nano-composite films. We suggest that this is a result of disorder that occurs in these mixed oxide immiscible systems, meaning large area devices can be reliably created through spontaneous interface grading. We expect that a better understanding of how to control the film-substrate interface in pinning-enhanced YBa$_2$Cu$_3$O$_{7-\delta}$ will permit further improvement of these diode-like properties. Further, the existence of this diode effect means that it is necessary to measure both in-plane peaks and avoid commutating measurements of voltage when studying the angular critical current dependence of such nano-composite films.

The authors would like to thank J. R. Clem and M. G. Blamire for helpful discussions. This work was supported by the Engineering and Physical Science Research Council [grant numbers EP/C011554/1, EP/C011546/1] and the EU Marie Curie Excellence programme [grant number MEXT-CT-2004-014156].


\begin{thebibliography}{28}
\expandafter\ifx\csname natexlab\endcsname\relax\fi
\expandafter\ifx\csname bibnamefont\endcsname\relax
  \def\bibnamefont#1{#1}\fi
\expandafter\ifx\csname bibfnamefont\endcsname\relax
  \def\bibfnamefont#1{#1}\fi
\expandafter\ifx\csname citenamefont\endcsname\relax
  \def\citenamefont#1{#1}\fi
\expandafter\ifx\csname url\endcsname\relax
  \def\url#1{\texttt{#1}}\fi
\expandafter\ifx\csname urlprefix\endcsname\relax\fi
\providecommand{\bibinfo}[2]{#2}
\providecommand{\eprint}[2][]{\url{#2}}

\bibitem[{\citenamefont{Krasnov et~al.}(1997)\citenamefont{Krasnov, Oboznov,
  and Pedersen}}]{Krasnov1997}
\bibinfo{author}{\bibfnamefont{V.~M.} \bibnamefont{Krasnov}},
  \bibinfo{author}{\bibfnamefont{V.~A.} \bibnamefont{Oboznov}},
  \bibnamefont{and} \bibinfo{author}{\bibfnamefont{N.~F.}
  \bibnamefont{Pedersen}}, \bibinfo{journal}{Physical Review B}
  \textbf{\bibinfo{volume}{55}}, \bibinfo{pages}{14486} (\bibinfo{year}{1997}).

\bibitem[{\citenamefont{Wordenweber et~al.}(2004)\citenamefont{Wordenweber,
  Dymashevski, and Misko}}]{Wordenweber2004}
\bibinfo{author}{\bibfnamefont{R.}~\bibnamefont{Wordenweber}},
  \bibinfo{author}{\bibfnamefont{P.}~\bibnamefont{Dymashevski}},
  \bibnamefont{and} \bibinfo{author}{\bibfnamefont{V.~R.} \bibnamefont{Misko}},
  \bibinfo{journal}{Physical Review B} \textbf{\bibinfo{volume}{69}},
  \bibinfo{pages}{184504} (\bibinfo{year}{2004}).

\bibitem[{\citenamefont{Lu et~al.}(2007)\citenamefont{Lu, Reichhardt, and
  Reichhardt}}]{Lu2007}
\bibinfo{author}{\bibfnamefont{Q.~M.} \bibnamefont{Lu}},
  \bibinfo{author}{\bibfnamefont{C.~J.~O.} \bibnamefont{Reichhardt}},
  \bibnamefont{and}
  \bibinfo{author}{\bibfnamefont{C.}~\bibnamefont{Reichhardt}},
  \bibinfo{journal}{Physical Review B} \textbf{\bibinfo{volume}{75}},
  \bibinfo{pages}{054502} (\bibinfo{year}{2007}).

\bibitem[{\citenamefont{Shalom and Pastoriza}(2005)}]{Shalom2005}
\bibinfo{author}{\bibfnamefont{D.~E.} \bibnamefont{Shalom}} \bibnamefont{and}
  \bibinfo{author}{\bibfnamefont{H.}~\bibnamefont{Pastoriza}},
  \bibinfo{journal}{Physical Review Letters} \textbf{\bibinfo{volume}{94}},
  \bibinfo{pages}{177001} (\bibinfo{year}{2005}).

\bibitem[{\citenamefont{Van~de Vondel et~al.}(2005)\citenamefont{Van~de Vondel,
  Silva, Zhu, Morelle, and Moshchalkov}}]{VandeVondel2005}
\bibinfo{author}{\bibfnamefont{J.}~\bibnamefont{Van~de Vondel}},
  \bibinfo{author}{\bibfnamefont{C.~C.~D.} \bibnamefont{Silva}},
  \bibinfo{author}{\bibfnamefont{B.~Y.} \bibnamefont{Zhu}},
  \bibinfo{author}{\bibfnamefont{M.}~\bibnamefont{Morelle}}, \bibnamefont{and}
  \bibinfo{author}{\bibfnamefont{V.~V.} \bibnamefont{Moshchalkov}},
  \bibinfo{journal}{Physical Review Letters} \textbf{\bibinfo{volume}{94}},
  \bibinfo{pages}{057003} (\bibinfo{year}{2005}).

\bibitem[{\citenamefont{Yu et~al.}(2007)\citenamefont{Yu, Heitmann, Song,
  Defeo, Plourde, Hesselberth, and Kes}}]{Yu2007}
\bibinfo{author}{\bibfnamefont{K.}~\bibnamefont{Yu}},
  \bibinfo{author}{\bibfnamefont{T.~W.} \bibnamefont{Heitmann}},
  \bibinfo{author}{\bibfnamefont{C.}~\bibnamefont{Song}},
  \bibinfo{author}{\bibfnamefont{M.~P.} \bibnamefont{Defeo}},
  \bibinfo{author}{\bibfnamefont{B.~L.~T.} \bibnamefont{Plourde}},
  \bibinfo{author}{\bibfnamefont{M.~B.~S.} \bibnamefont{Hesselberth}},
  \bibnamefont{and} \bibinfo{author}{\bibfnamefont{P.~H.} \bibnamefont{Kes}},
  \bibinfo{journal}{Physical Review B} \textbf{\bibinfo{volume}{76}},
  \bibinfo{pages}{220507} (\bibinfo{year}{2007}).

\bibitem[{\citenamefont{Cole et~al.}(2006)\citenamefont{Cole, Bending,
  Savel'ev, Grigorenko, Tamegai, and Nori}}]{Cole2006}
\bibinfo{author}{\bibfnamefont{D.}~\bibnamefont{Cole}},
  \bibinfo{author}{\bibfnamefont{S.}~\bibnamefont{Bending}},
  \bibinfo{author}{\bibfnamefont{S.}~\bibnamefont{Savel'ev}},
  \bibinfo{author}{\bibfnamefont{A.}~\bibnamefont{Grigorenko}},
  \bibinfo{author}{\bibfnamefont{T.}~\bibnamefont{Tamegai}}, \bibnamefont{and}
  \bibinfo{author}{\bibfnamefont{F.}~\bibnamefont{Nori}},
  \bibinfo{journal}{Nature Materials} \textbf{\bibinfo{volume}{5}},
  \bibinfo{pages}{305} (\bibinfo{year}{2006}).

\bibitem[{\citenamefont{Broussard and Geballe}(1988)}]{BROUSSARD1988}
\bibinfo{author}{\bibfnamefont{P.~R.} \bibnamefont{Broussard}}
  \bibnamefont{and} \bibinfo{author}{\bibfnamefont{T.~H.}
  \bibnamefont{Geballe}}, \bibinfo{journal}{Physical Review B}
  \textbf{\bibinfo{volume}{37}}, \bibinfo{pages}{68} (\bibinfo{year}{1988}).

\bibitem[{\citenamefont{Jiang et~al.}(1994)\citenamefont{Jiang, Connonlly,
  Hagen, and Lobb}}]{JIANG1994}
\bibinfo{author}{\bibfnamefont{X.~G.} \bibnamefont{Jiang}},
  \bibinfo{author}{\bibfnamefont{P.~J.} \bibnamefont{Connonlly}},
  \bibinfo{author}{\bibfnamefont{S.~J.} \bibnamefont{Hagen}}, \bibnamefont{and}
  \bibinfo{author}{\bibfnamefont{C.~J.} \bibnamefont{Lobb}},
  \bibinfo{journal}{Physical Review B} \textbf{\bibinfo{volume}{49}},
  \bibinfo{pages}{9244} (\bibinfo{year}{1994}).

\bibitem[{\citenamefont{Jiang and Lobb}(1995)}]{JIANG1995}
\bibinfo{author}{\bibfnamefont{X.~G.} \bibnamefont{Jiang}} \bibnamefont{and}
  \bibinfo{author}{\bibfnamefont{C.~J.} \bibnamefont{Lobb}},
  \bibinfo{journal}{Journal Of Low Temperature Physics}
  \textbf{\bibinfo{volume}{100}}, \bibinfo{pages}{515} (\bibinfo{year}{1995}).

\bibitem[{\citenamefont{Harrington et~al.}(2009)\citenamefont{Harrington,
  Durrell, Maiorov, Wang, Wimbush, Kursumovic, Lee, and
  MacManus-Driscoll}}]{Harrington2009}
\bibinfo{author}{\bibfnamefont{S.~A.} \bibnamefont{Harrington}},
  \bibinfo{author}{\bibfnamefont{J.~H.} \bibnamefont{Durrell}},
  \bibinfo{author}{\bibfnamefont{B.}~\bibnamefont{Maiorov}},
  \bibinfo{author}{\bibfnamefont{H.}~\bibnamefont{Wang}},
  \bibinfo{author}{\bibfnamefont{S.~C.} \bibnamefont{Wimbush}},
  \bibinfo{author}{\bibfnamefont{A.}~\bibnamefont{Kursumovic}},
  \bibinfo{author}{\bibfnamefont{J.~H.} \bibnamefont{Lee}}, \bibnamefont{and}
  \bibinfo{author}{\bibfnamefont{J.~L.} \bibnamefont{MacManus-Driscoll}},
  \bibinfo{journal}{Superconductor Science \& Technology}
  \textbf{\bibinfo{volume}{22}}, \bibinfo{pages}{022001}
  (\bibinfo{year}{2009}).

\bibitem[{\citenamefont{Maiorov et~al.}(2009)\citenamefont{Maiorov, Baily,
  Zhou, Ugurlu, Kennison, Dowden, Holesinger, Foltyn, and
  Civale}}]{B.Maiorov2009}
\bibinfo{author}{\bibfnamefont{B.}~\bibnamefont{Maiorov}},
  \bibinfo{author}{\bibfnamefont{S.~A.} \bibnamefont{Baily}},
  \bibinfo{author}{\bibfnamefont{H.}~\bibnamefont{Zhou}},
  \bibinfo{author}{\bibfnamefont{O.}~\bibnamefont{Ugurlu}},
  \bibinfo{author}{\bibfnamefont{J.~A.} \bibnamefont{Kennison}},
  \bibinfo{author}{\bibfnamefont{P.~C.} \bibnamefont{Dowden}},
  \bibinfo{author}{\bibfnamefont{T.~G.} \bibnamefont{Holesinger}},
  \bibinfo{author}{\bibfnamefont{S.~R.} \bibnamefont{Foltyn}},
  \bibnamefont{and} \bibinfo{author}{\bibfnamefont{L.}~\bibnamefont{Civale}},
  \bibinfo{journal}{Nature Materials} \textbf{\bibinfo{volume}{8}}, \bibinfo{pages}{398-404} (\bibinfo{year}{2009}).

\bibitem[{\citenamefont{Mele et~al.}(2008)\citenamefont{Mele, Matsumoto,
  Ichinose, Mukaida, Yoshida, Horii, and Kita}}]{Mele2008}
\bibinfo{author}{\bibfnamefont{P.}~\bibnamefont{Mele}},
  \bibinfo{author}{\bibfnamefont{K.}~\bibnamefont{Matsumoto}},
  \bibinfo{author}{\bibfnamefont{A.}~\bibnamefont{Ichinose}},
  \bibinfo{author}{\bibfnamefont{M.}~\bibnamefont{Mukaida}},
  \bibinfo{author}{\bibfnamefont{Y.}~\bibnamefont{Yoshida}},
  \bibinfo{author}{\bibfnamefont{S.}~\bibnamefont{Horii}}, \bibnamefont{and}
  \bibinfo{author}{\bibfnamefont{R.}~\bibnamefont{Kita}},
  \bibinfo{journal}{Superconductor Science \& Technology}
  \textbf{\bibinfo{volume}{21}}, \bibinfo{pages}{125017}
  (\bibinfo{year}{2008}).

\bibitem[{\citenamefont{Roas et~al.}(1990)\citenamefont{Roas, Schultz, and
  Saemannishenko}}]{roa90}
\bibinfo{author}{\bibfnamefont{B.}~\bibnamefont{Roas}},
  \bibinfo{author}{\bibfnamefont{L.}~\bibnamefont{Schultz}}, \bibnamefont{and}
  \bibinfo{author}{\bibfnamefont{G.}~\bibnamefont{Saemannishenko}},
  \bibinfo{journal}{Phys. Rev. Lett.} \textbf{\bibinfo{volume}{64}},
  \bibinfo{pages}{479} (\bibinfo{year}{1990}).

\bibitem[{\citenamefont{Civale et~al.}(2004)\citenamefont{Civale, Maiorov,
  Serquis, Willis, Coulter, Wang, Jia, Arendt, MacManus-Driscoll, Maley
  et~al.}}]{civ04}
\bibinfo{author}{\bibfnamefont{L.}~\bibnamefont{Civale}},
  \bibinfo{author}{\bibfnamefont{B.}~\bibnamefont{Maiorov}},
  \bibinfo{author}{\bibfnamefont{A.}~\bibnamefont{Serquis}},
  \bibinfo{author}{\bibfnamefont{J.~O.} \bibnamefont{Willis}},
  \bibinfo{author}{\bibfnamefont{J.~Y.} \bibnamefont{Coulter}},
  \bibinfo{author}{\bibfnamefont{H.}~\bibnamefont{Wang}},
  \bibinfo{author}{\bibfnamefont{Q.~X.} \bibnamefont{Jia}},
  \bibinfo{author}{\bibfnamefont{P.~N.} \bibnamefont{Arendt}},
  \bibinfo{author}{\bibfnamefont{J.~L.} \bibnamefont{MacManus-Driscoll}},
  \bibinfo{author}{\bibfnamefont{M.~P.} \bibnamefont{Maley}},
  \bibnamefont{and} \bibinfo{author}{\bibfnamefont{S.~R.} \bibnamefont{Foltyn}}
  , \bibinfo{journal}{Applied Physics Letters}
  \textbf{\bibinfo{volume}{84}}, \bibinfo{pages}{2121} (\bibinfo{year}{2004}).

\bibitem[{\citenamefont{Nishizaki et~al.}(1991)\citenamefont{Nishizaki, Aomine,
  Fujii, Yamamoto, Yoshii, Terashima, and Bando}}]{nis91}
\bibinfo{author}{\bibfnamefont{T.}~\bibnamefont{Nishizaki}},
  \bibinfo{author}{\bibfnamefont{T.}~\bibnamefont{Aomine}},
  \bibinfo{author}{\bibfnamefont{I.}~\bibnamefont{Fujii}},
  \bibinfo{author}{\bibfnamefont{K.}~\bibnamefont{Yamamoto}},
  \bibinfo{author}{\bibfnamefont{S.}~\bibnamefont{Yoshii}},
  \bibinfo{author}{\bibfnamefont{T.}~\bibnamefont{Terashima}},
  \bibnamefont{and} \bibinfo{author}{\bibfnamefont{Y.}~\bibnamefont{Bando}},
  \bibinfo{journal}{Physica C} \textbf{\bibinfo{volume}{185}},
  \bibinfo{pages}{2259} (\bibinfo{year}{1991}).

\bibitem[{\citenamefont{Tachiki and Takahashi}(1989)}]{tac89}
\bibinfo{author}{\bibfnamefont{M.}~\bibnamefont{Tachiki}} \bibnamefont{and}
  \bibinfo{author}{\bibfnamefont{S.}~\bibnamefont{Takahashi}},
  \bibinfo{journal}{Solid State Communications} \textbf{\bibinfo{volume}{70}},
  \bibinfo{pages}{291} (\bibinfo{year}{1989}).

\bibitem[{\citenamefont{Maiorov et~al.}(2007)\citenamefont{Maiorov, Jia, Zhou,
  Wang, Li, Kursumovic, MacManus-Driscoll, Haugan, Barnes, Foltyn
  et~al.}}]{Maiorov2007}
\bibinfo{author}{\bibfnamefont{B.}~\bibnamefont{Maiorov}},
  \bibinfo{author}{\bibfnamefont{Q.~X.} \bibnamefont{Jia}},
  \bibinfo{author}{\bibfnamefont{H.}~\bibnamefont{Zhou}},
  \bibinfo{author}{\bibfnamefont{H.}~\bibnamefont{Wang}},
  \bibinfo{author}{\bibfnamefont{Y.}~\bibnamefont{Li}},
  \bibinfo{author}{\bibfnamefont{A.}~\bibnamefont{Kursumovic}},
  \bibinfo{author}{\bibfnamefont{J.~L.} \bibnamefont{MacManus-Driscoll}},
  \bibinfo{author}{\bibfnamefont{T.~J.} \bibnamefont{Haugan}},
  \bibinfo{author}{\bibfnamefont{P.~N.} \bibnamefont{Barnes}},
  \bibinfo{author}{\bibfnamefont{S.~R.} \bibnamefont{Foltyn}},
  \bibnamefont{and} \bibinfo{author}{\bibfnamefont{L.}~\bibnamefont{Civale}},
   \bibinfo{journal}{IEEE Transactions on Applied
  Superconductivity} \textbf{\bibinfo{volume}{17}}, \bibinfo{pages}{3697}
  (\bibinfo{year}{2007}).

\bibitem[{\citenamefont{Berghuis et~al.}(1996)\citenamefont{Berghuis,
  Baudenbacher, Herzog, Somekh, Campbell, Evetts, and Wirth}}]{ber96_3}
\bibinfo{author}{\bibfnamefont{P.}~\bibnamefont{Berghuis}},
  \bibinfo{author}{\bibfnamefont{F.}~\bibnamefont{Baudenbacher}},
  \bibinfo{author}{\bibfnamefont{R.}~\bibnamefont{Herzog}},
  \bibinfo{author}{\bibfnamefont{R.}~\bibnamefont{Somekh}},
  \bibinfo{author}{\bibfnamefont{A.~M.} \bibnamefont{Campbell}},
  \bibinfo{author}{\bibfnamefont{J.~E.} \bibnamefont{Evetts}},
  \bibnamefont{and} \bibinfo{author}{\bibfnamefont{G.}~\bibnamefont{Wirth}}, in
  \emph{\bibinfo{booktitle}{Proc. 8th IWCC in Superconductors}}, edited by
  \bibinfo{editor}{\bibfnamefont{T.~M.} \bibnamefont{Matsushita}}
  \bibnamefont{and} \bibinfo{editor}{\bibfnamefont{K.}~\bibnamefont{Yamafuji}}
  (\bibinfo{publisher}{World Scientific, Singapore}, \bibinfo{year}{1996}), p.
  \bibinfo{pages}{247}.

\bibitem[{\citenamefont{Campbell and Evetts}(1972)}]{cam72}
\bibinfo{author}{\bibfnamefont{A.}~\bibnamefont{Campbell}} \bibnamefont{and}
  \bibinfo{author}{\bibfnamefont{J.}~\bibnamefont{Evetts}},
  \bibinfo{journal}{Advances in Physics} \textbf{\bibinfo{volume}{50}},
  \bibinfo{pages}{1249} (\bibinfo{year}{1972}).

\bibitem[{\citenamefont{Bean and Livingston}(1964)}]{Bean1964}
\bibinfo{author}{\bibfnamefont{C.~P.} \bibnamefont{Bean}} \bibnamefont{and}
  \bibinfo{author}{\bibfnamefont{J.~D.} \bibnamefont{Livingston}},
  \bibinfo{journal}{Physics Review Letters} \textbf{\bibinfo{volume}{12}},
  \bibinfo{pages}{14} (\bibinfo{year}{1964}).

\bibitem[{\citenamefont{Clem}(1974)}]{clem74}
\bibinfo{author}{\bibfnamefont{J.~R.} \bibnamefont{Clem}}, in
  \emph{\bibinfo{booktitle}{Low Temperature Physics - LT13}}, edited by
  \bibinfo{editor}{\bibfnamefont{K.~D.} \bibnamefont{Timmerhaus}},
  \bibinfo{editor}{\bibfnamefont{W.~J.} \bibnamefont{O'Sullivan}},
  \bibnamefont{and} \bibinfo{editor}{\bibfnamefont{E.~F.} \bibnamefont{Hammel}}
  (\bibinfo{year}{1974}), vol.~\bibinfo{volume}{3}, pp.
  \bibinfo{pages}{102--106}.

\bibitem[{\citenamefont{Vodolazov and Peeters}(2005)}]{Vodolazov2005}
\bibinfo{author}{\bibfnamefont{D.~Y.} \bibnamefont{Vodolazov}}
  \bibnamefont{and} \bibinfo{author}{\bibfnamefont{F.~M.}
  \bibnamefont{Peeters}}, \bibinfo{journal}{Physical Review B}
  \textbf{\bibinfo{volume}{72}}, \bibinfo{pages}{172508}
  (\bibinfo{year}{2005}).

\bibitem[{\citenamefont{Kumar et~al.}(1994)\citenamefont{Kumar, Blamire, Doyle,
  Campbell, and Evetts}}]{KUMAR1994}
\bibinfo{author}{\bibfnamefont{D.~D.} \bibnamefont{Kumar}},
  \bibinfo{author}{\bibfnamefont{M.~G.} \bibnamefont{Blamire}},
  \bibinfo{author}{\bibfnamefont{R.}~\bibnamefont{Doyle}},
  \bibinfo{author}{\bibfnamefont{A.~M.} \bibnamefont{Campbell}},
  \bibnamefont{and} \bibinfo{author}{\bibfnamefont{J.~E.}
  \bibnamefont{Evetts}}, \bibinfo{journal}{Journal Of Applied Physics}
  \textbf{\bibinfo{volume}{76}}, \bibinfo{pages}{2361} (\bibinfo{year}{1994}).

\bibitem[{\citenamefont{Herzog and Evetts}(1994)}]{her94}
\bibinfo{author}{\bibfnamefont{R.}~\bibnamefont{Herzog}} \bibnamefont{and}
  \bibinfo{author}{\bibfnamefont{J.~E.} \bibnamefont{Evetts}},
  \bibinfo{journal}{Review Of Scientific Instruments}
  \textbf{\bibinfo{volume}{65}}, \bibinfo{pages}{3574} (\bibinfo{year}{1994}).

\bibitem[{\citenamefont{Durrell et~al.}(2003)\citenamefont{Durrell, Hogg,
  Kahlmann, Barber, Blamire, and Evetts}}]{durrgb03}
\bibinfo{author}{\bibfnamefont{J.~H.} \bibnamefont{Durrell}},
  \bibinfo{author}{\bibfnamefont{M.~J.} \bibnamefont{Hogg}},
  \bibinfo{author}{\bibfnamefont{F.}~\bibnamefont{Kahlmann}},
  \bibinfo{author}{\bibfnamefont{Z.~H.} \bibnamefont{Barber}},
  \bibinfo{author}{\bibfnamefont{M.~G.} \bibnamefont{Blamire}},
  \bibnamefont{and} \bibinfo{author}{\bibfnamefont{J.~E.}
  \bibnamefont{Evetts}}, \bibinfo{journal}{Physical Review Letters}
  \textbf{\bibinfo{volume}{90}}, \bibinfo{pages}{247006}
  (\bibinfo{year}{2003}).

\bibitem[{\citenamefont{Durrell et~al.}(2004)\citenamefont{Durrell, Burnell,
  Barber, Blamire, and Evetts}}]{dur04}
\bibinfo{author}{\bibfnamefont{J.~H.} \bibnamefont{Durrell}},
  \bibinfo{author}{\bibfnamefont{G.}~\bibnamefont{Burnell}},
  \bibinfo{author}{\bibfnamefont{Z.~H.} \bibnamefont{Barber}},
  \bibinfo{author}{\bibfnamefont{M.~G.} \bibnamefont{Blamire}},
  \bibnamefont{and} \bibinfo{author}{\bibfnamefont{J.~E.}
  \bibnamefont{Evetts}}, \bibinfo{journal}{Physical Review B}
  \textbf{\bibinfo{volume}{70}}, \bibinfo{pages}{214508}
  (\bibinfo{year}{2004}).

\bibitem[{\citenamefont{Goyal et~al.}(2005)\citenamefont{Goyal, Kang, Leonard,
  Martin, Gapud, Varela, Paranthaman, Ijaduola, Specht, Thompson
  et~al.}}]{Goyal2005a}
\bibinfo{author}{\bibfnamefont{A.}~\bibnamefont{Goyal}},
  \bibinfo{author}{\bibfnamefont{S.}~\bibnamefont{Kang}},
  \bibinfo{author}{\bibfnamefont{K.~J.} \bibnamefont{Leonard}},
  \bibinfo{author}{\bibfnamefont{P.~M.} \bibnamefont{Martin}},
  \bibinfo{author}{\bibfnamefont{A.~A.} \bibnamefont{Gapud}},
  \bibinfo{author}{\bibfnamefont{M.}~\bibnamefont{Varela}},
  \bibinfo{author}{\bibfnamefont{M.}~\bibnamefont{Paranthaman}},
  \bibinfo{author}{\bibfnamefont{A.~O.} \bibnamefont{Ijaduola}},
  \bibinfo{author}{\bibfnamefont{E.~D.} \bibnamefont{Specht}},
  \bibinfo{author}{\bibfnamefont{J.~R.} \bibnamefont{Thompson}},
  \bibinfo{author}{\bibfnamefont{D.~K.} \bibnamefont{Christen}},
  \bibinfo{author}{\bibfnamefont{S.~J.} \bibnamefont{Pennycook}}
  \bibnamefont{and} \bibinfo{author}{\bibfnamefont{F.~A.} \bibnamefont{List}},
  \bibinfo{journal}{Superconductor Science \& Technology}
  \textbf{\bibinfo{volume}{18}}, \bibinfo{pages}{1533} (\bibinfo{year}{2005}).

\end{thebibliography}
\end{document}